\documentclass[letterpaper]{article} % DO NOT CHANGE THIS
\usepackage{aaai25}  % DO NOT CHANGE THIS
\usepackage{times}  % DO NOT CHANGE THIS
\usepackage{helvet}  % DO NOT CHANGE THIS
\usepackage{courier}  % DO NOT CHANGE THIS
\usepackage[hyphens]{url}  % DO NOT CHANGE THIS
\usepackage{graphicx} % DO NOT CHANGE THIS
\usepackage{natbib}  % DO NOT CHANGE THIS AND DO NOT ADD ANY OPTIONS TO IT
\usepackage{caption} % DO NOT CHANGE THIS AND DO NOT ADD ANY OPTIONS TO IT
\usepackage{algpseudocode}
\usepackage{algorithm}
\usepackage{multirow}
\usepackage{booktabs}
\usepackage{xspace}
\usepackage{pifont}
\usepackage{amsfonts}
\usepackage{amsmath}
\usepackage{xcolor}
\usepackage{makecell}

\usepackage{mdframed}
\usepackage{lipsum} % for dummy text

\newmdenv[
  topline=false,
  bottomline=false,
  rightline=false,
  skipabove=\baselineskip,
  skipbelow=\baselineskip,
  linecolor=black,
  linewidth=1pt
]{myuline}

\makeatletter
\DeclareRobustCommand\onedot{\futurelet\@let@token\@onedot}
\def\@onedot{\ifx\@let@token.\else.\null\fi\xspace}
\def\eg{\emph{e.g}\onedot} 
\def\ie{\emph{i.e}\onedot}

\makeatother

\newcommand{\tool}{{\sc \texttt{Dreamark}}\xspace}

\usepackage{newfloat}
\usepackage{listings}
\DeclareCaptionStyle{ruled}{labelfont=normalfont,labelsep=colon,strut=off} % DO NOT CHANGE THIS
\floatstyle{ruled}
\newfloat{listing}{tb}{lst}{}
\floatname{listing}{Listing}
\title{DreaMark: Rooting Watermark in Score Distillation Sampling Generated Neural Radiance Fields}
\author{
    Xingyu Zhu\textsuperscript{\rm 1,\rm 2},
    Xiapu Luo\textsuperscript{\rm 2},
    Xuetao Wei\textsuperscript{\rm 1}\thanks{Corresponding Author.}
}
\affiliations{
    \textsuperscript{\rm 1}Department of Computer Science and Engineering, Southern University of Science and Technology, China \\
    \textsuperscript{\rm 2}Department of Computing, Hong Kong Polytechnic University, Hong Kong\\
    12150086@mail.sustech.edu.cn, csxluo@comp.polyu.edu.hk, weixt@sustech.edu.cn
}

\usepackage{bibentry}
\begin{document}

\maketitle

\begin{abstract}
Recent advancements in text-to-3D generation can generate neural radiance fields (NeRFs) with score distillation sampling, enabling 3D asset creation without real-world data capture. With the rapid advancement in NeRF generation quality, protecting the copyright of the generated NeRF has become increasingly important. While prior works can watermark NeRFs in a post-generation way, they suffer from two vulnerabilities. First, a delay lies between NeRF generation and watermarking because the secret message is embedded into the NeRF model post-generation through fine-tuning. Second, generating a non-watermarked NeRF as an intermediate creates a potential vulnerability for theft. To address both issues, we propose \tool to embed a secret message by backdooring the NeRF during NeRF generation. In detail, we first pre-train a watermark decoder. Then, \tool generates backdoored NeRFs in a way that the target secret message can be verified by the pre-trained watermark decoder on an arbitrary trigger viewport. We evaluate the generation quality and watermark robustness against image- and model-level attacks. Extensive experiments show that the watermarking process will not degrade the generation quality, and the watermark achieves 90+\% accuracy among both image-level attacks (\eg, Gaussian noise) and model-level attacks (\eg, pruning attack).
\end{abstract}


\section{Introduction}

Digital 3D content has become indispensable in Metaverse and virtual and augmented reality, enabling visualization, comprehension, and interaction with complex scenes that represent our real lives. Recent progress in 3D content generation \cite{poole2022dreamfusion,lin2023magic3d,wang2024prolificdreamer,liang2024luciddreamer} can generate high-quality 3D assets that need a lot of time, computational resources, and skilled expertise. Therefore, protecting the ownership of generated 3D content has become more critical.

We focus on Text-to-3D generation \cite{poole2022dreamfusion,lin2023magic3d,wang2024prolificdreamer,liang2024luciddreamer} and the neural radiance field (NeRF) \cite{mildenhall2021nerf, muller2022instant}, which have emerged into the spotlight in 3D content modeling. Current trending 3D generation algorithms generate 3D representations such as meshes and NeRFs. This paper focuses on NeRF generation since NeRF can represent 3D models more compactly. Given a textual description, recent text-to-3D methods generate NeRFs by distilling pre-trained diffusion models, such as Stable Diffusion \cite{rombach2022high}. This remarkable progress is grounded in the use of Score Distillation Sampling (SDS). With SDS, NeRF training can be conducted without realistic images. Thus, the research question we address in this paper is: \textit{how to protect the score distillation sampling generated neural radiance fields?}

\begin{figure}[t]
    \centering
    \includegraphics[width=\linewidth]{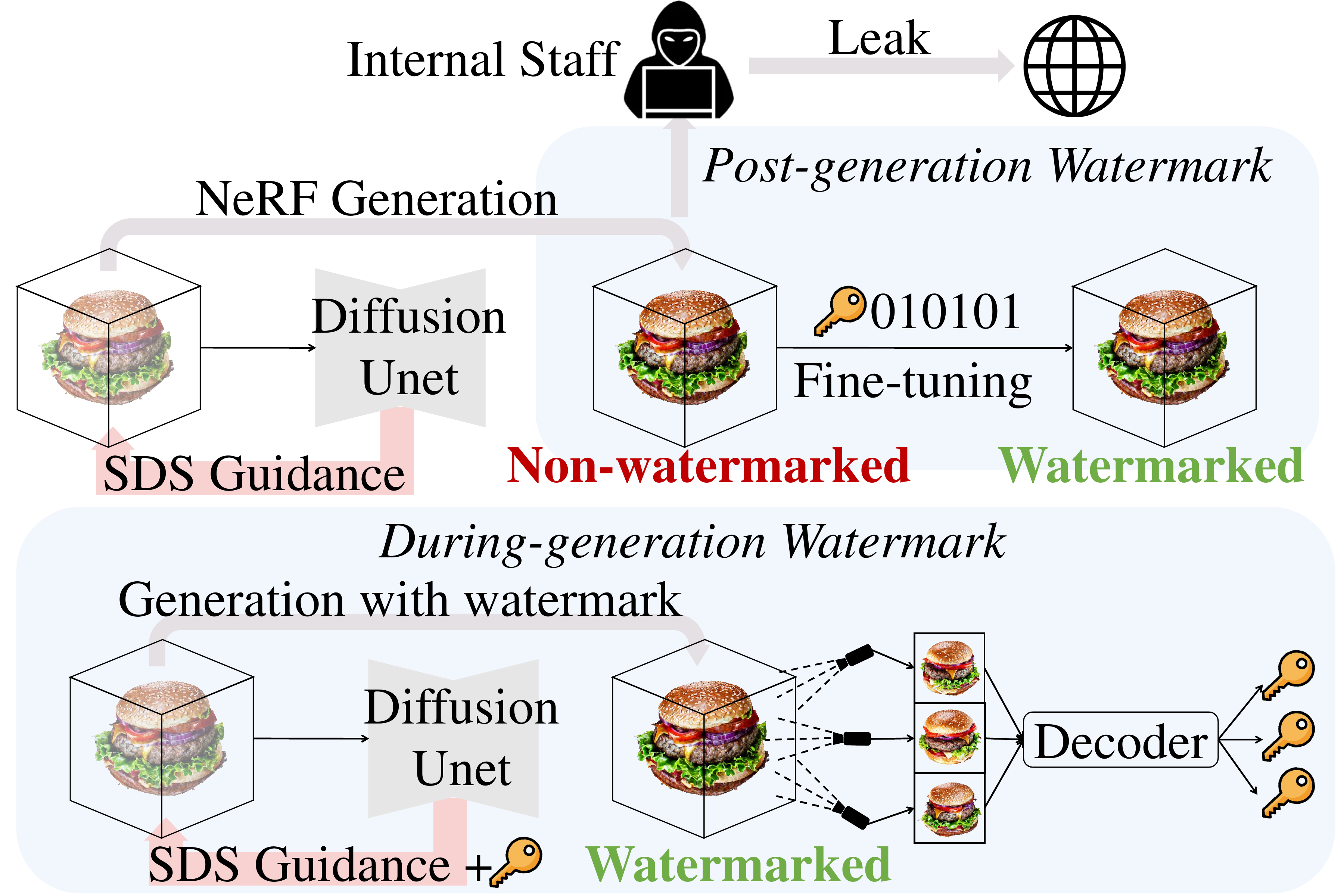}
    \caption{Attack scenario. If company-generated content is considered company property, internal staff could steal non-watermarked intermediates in the post-generation pipeline (top row). However, such intermediates do not exist in the during-generation pipeline (bottom row).}
    \label{fig:motivation_egs}
\end{figure}

One way to protect the generated NeRF is to apply post-generation watermarking methods, such as CopyRNeRF \cite{luo2023copyrnerf} and WateNeRF \cite{jang2024waterf}, to watermark NeRF after it is generated. However, these methods exhibit two problems. First, post-generation methods pose a risk of data leakage. As shown in Figure \ref{fig:motivation_egs}, since non-watermarked intermediates are generated in the post-generation pipeline, a malicious user could leak the non-watermarked version of the generated content. Second, CopyRNeRF increases the watermarking expense since it requires an additional message feature field in the NeRF structure. Integrating CopyRNeRF with an arbitrary text-to-NeRF pipeline requires additional modifications to the NeRF structure. Recognizing these limitations of previous work, can we conduct a \textit{during-generation} watermarking without modifying the NeRF structure?

We propose \tool, the first \textit{during-generation} text-to-3D watermarking method, which is gracefully combined with score distillation sampling to generate high-quality and watermarked NeRF. Different from \textit{post-generation} NeRF watermarking method, \tool directly generates watermarked NeRF without changing NeRF architecture, increasing the flexibility for future development on 3D generation. Our method is inspired by black-box model watermarking methods \cite{adi2018turning, zhang2018protecting, jia2021entangled, le2020adversarial, chen2019blackmarks, szyller2021dawn} which watermark a deep neural network by injecting backdoors. To inject backdoors in NeRF during generation, we first generate a trigger view set dependent only on the given secret message. Then, we conduct score distillation sampling in a way that the secret message can be extracted from images rendered from arbitrary trigger viewports. To extract the secret message from the rendered image, we use a pre-trained watermark decoder from HiDDeN \cite{zhu2018hidden}. All NeRF generated by \tool can be verified as watermarked by such a unique decoder.

Two critical evaluation metrics for watermarking algorithms are invisibility and robustness. For robustness, we evaluate bit accuracy under multiple image transformations, such as Gaussian noise, before images rendered from trigger viewports are fed into the watermark decoder. For invisibility, there is no such the ``original NeRF" since we root watermarks during a generation task, so \tool cannot be evaluated by Peak Signal-to-Noise Ratio (PSNR) as is done in \textit{post-generation} methods. However, we can still evaluate the invisibility by evaluating the generation quality as previous 2D watermarking tasks \cite{wen2024tree, yang2024gaussian}, where they use CLIP Score \cite{radford2021clip} to show the generation quality. Extensive experiments show that \tool successfully embeds the watermark in a \textit{during-generation} way and maintains robustness under multiple image transformations without degrading the generation quality. In summary, our contributions are as follows:
\begin{itemize}
    \item To the best of our knowledge, we propose \tool, the first during-generation 3D watermarking method, which eliminates the delay between NeRF generation and watermarking, ensuring that no non-watermarked version of the NeRF is ever produced, thereby preventing NeRF theft.
    
    \item The key novelty of our \tool is that it watermarks NeRF by injecting backdoors during score distillation sampling, such that the secret message can be extracted from images rendered from arbitrary trigger viewport.
  
    \item Extensive experiments show that the embedded watermark achieves 90+\% bit accuracy against multiple image transformations, and the watermarking process does not degrade the generation quality.
\end{itemize}
\section{Related Work}

\subsection{Text-to-3D Content Generation}

One category of text-to-3D generation starts from DreamField \cite{jain2022zero}, which trains NeRF with CLIP \cite{radford2021clip} guidance to achieve text-to-3D distillation. However, the generated content is unsatisfactory due to the weak supervision from CLIP loss. Hence, our work will not consider watermarking CLIP-guided 3D content generation. Another category starts from Dreamfusion \cite{poole2022dreamfusion}, which pioneerly introduces Score Distillation Sampling (SDS) to optimize NeRF by distilling a pre-trained text-to-image diffusion model. This motivates a great number of following works to propose critical incremental. These works improve the quality of generation in various ways. For example, Fantasia3D \cite{chen2023fantasia3d}, Magic3D \cite{lin2023magic3d}, Latent-nerf \cite{metzer2023latent}, DreamGaussian \cite{tang2023dreamgaussian} and GaussianDreamer \cite{yi2023gaussiandreamer} improve the visual quality of generated content by changing 3D representations or improving NeRF structure. MVDream \cite{shi2023mvdream} focuses on addressing Janus problems by fine-tuning the pre-trained diffusion model to make it 3D aware. However, SDS guidance still suffers from over-saturation problems, as is shown in Magic3D \cite{lin2023magic3d}, Dreamfusion \cite{poole2022dreamfusion}, and AvatarVerse \cite{zhang2024avatarverse}. The other, like ProlificDreamer \cite{wang2024prolificdreamer} and LucidDreamer \cite{liang2024luciddreamer}, focus on improving SDS itself. For example, LucidDreamer uncovers the reason for the overly-smoothed problem that SDS guides the generation process towards an averaged pseudo-ground-truth and proposes ISM to relieve such a problem. ProlificDreamer proposes VSD guidance instead of SDS guidance and shows that SDS is just a special case of VSD. Although extensive research has been proposed to improve text-to-3D generation, these works still require a much longer training stage, which makes it necessary to protect the copyright of generated content.

\subsection{Digital Watermarking}

Digital watermarking hides watermarks into multimedia for copyright protection or leakage source tracing. Various research works have been proposed to protect traditional multimedia content like 2D images and 3D meshes. Early works watermark images and meshes by embedding a secret message in either the least significant bits \cite{van1994digital} or the most significant bits \cite{tsai2020separable, jiang2017reversible, tsai2022integrating, peng2022semi, peng2021general} of image pixels and vertex coordinates. HiDDeN \cite{zhu2018hidden} and Deep3DMark \cite{zhu2024rethinking} have made substantial improvements using deep learning networks.

Recently, several watermarking methods have emerged in the NeRF domain. StegaNeRF \cite{li2023steganerf} designed a steganography algorithm that hides natural images in 3D scene representation. CopyRNeRF \cite{luo2023copyrnerf} protects the copyright of NeRF by verifying the secret message extracted from images rendered from the protected NeRF. WateNeRF \cite{jang2024waterf} further improves NeRF watermarking by hiding secret messages into the frequency domain of rendered images, increasing the robustness of the watermark. However, CopyRNeRF and WateNeRF are two \textit{post-generation} watermarking methods, \ie they watermark by fine-tuning a pre-trained NeRF. This poses a delay between the NeRF generation and watermarking. A malicious user could obtain the pre-trained NeRF before it is watermarked. Besides, CopyRNeRF requires additional changes in NeRF architecture. We would like the watermarking method to be architecture agnostic due to the fact that some text-to-3D generation methods, like Magic3D, Fantasia3D, and Latent-nerf, require specific NeRF architecture for visual quality improvement. To address these issues, we design an architecture-agnostic method that watermarks NeRF during generation.
\section{Preliminaries}

\textbf{NeRF.} NeRF \cite{mildenhall2021nerf} uses multilayer perceptrons (MLPs) $f_{\sigma}$ and $f_{\mathbf{c}}$ to map the 3D location $\mathbf{x}\in \mathbb{R}^3$ and viewing direction $\mathbf{d}\in\mathbb{R}^2$ to a color value $\mathbf{c}\in\mathbb{R}^3$ and a geometric value $\sigma \in \mathbb{R}^+$:
\begin{equation}
    \sigma, \mathbf{z} = f_\sigma(\gamma_\mathbf{x}(\mathbf{x})),
    \label{eq:sigma}
\end{equation}
\begin{equation}
    \mathbf{c}=f_\mathbf{c}(\mathbf{z}, \gamma_\mathbf{d}(\mathbf{d})),
    \label{eq:color}
\end{equation}
where $\gamma_{\mathbf{x}}, \gamma_{\mathbf{d}}$ are fixed encoding functions for location and viewing direction, respectively. The intermediate variable $\mathbf{z}$ is a feature output by the first MLP $f_\sigma$. To render a $H\times W$ image with the given viewport $\mathbf{p}$, the rendering process casts rays $\{r_i\}_{i=1}^{H\times W}$ from pixels and computes the weighted sum of the color $\mathbf{c}_j$ of the sampling points along each ray to composite the color of each pixel:
\begin{equation}
    \hat{\mathbf{C}}(r_i)=\sum_j T_j(1-\exp(-\sigma_j \delta_j))\mathbf{c}_j,
    \label{eq:ray_color}
\end{equation}
where $T_j=\prod_k^{j-1} \exp (-\sigma_k\delta_k)$, and $\delta_k$ is the distance between adjacent sample points. In later chapters, we use $\mathbf{g}(\theta, \mathbf{p})\in [0,1]^{H\times W\times 3}$ to represent the above rendering process, where $\theta$ represents parameters of a NeRF, and $\mathbf{g}$ takes viewport $\mathbf{p}$ as input and outputs a normalized image.

\textbf{Diffusion models.} A diffusion model \cite{song2020score, ho2020denoising, song2020denoising} involves a forward process $\{q_t\}_{t\in [0,1]}$ to gradually add noise to a data point $\mathbf{x}_0\sim q_0(\mathbf{x}_0)$ and a reverse process $\{p_t\}_{t\in[0,1]}$ to denoise/generate data. The forward process is defined by $q_t(\mathbf{x}_t|\mathbf{x}_0):=\mathcal{N}(\alpha_t\mathbf{x}_0,\sigma^2_t\mathbf{I})$ and $q_t(\mathbf{x}_t):=\int q_t(\mathbf{x}_t|\mathbf{x}_0)q_0(\mathbf{x}_0)d\mathbf{x}_0$, where $\alpha_t,\sigma_t>0$ are hyperparamaters; and the reverse process is defined by denoising from $p_1(\mathbf{x}_1):=\mathcal{N}(\mathbf{0}, \mathbf{I})$ with a parameterized noise prediction network $\mathbf{\epsilon}_\phi(\mathbf{x}_t, t)$ to predict the noise added to a clean data $\mathbf{x}_0$, which is trained by minimizing:
\begin{equation}
    %\mathcal{L}_{\text{Diff}}=\mathbb{E}_{\mathbf{x}_0\sim q_0(\mathbf{x}_0),t\sim \mathcal{U}(0,1),\mathbf{\epsilon}\sim \mathcal{N}(\mathbf{0},\mathbf{I})}\left[w(t)\lVert \epsilon_\phi(\alpha_t \mathbf{x}_0+\sigma_t\mathbf{\epsilon})-\epsilon\rVert _2^2\right]
    \mathcal{L}_{\text{Diff}}(\phi)=\mathbb{E}_{\mathbf{x}_0,t,\epsilon}\left[w(t)\lVert \epsilon_\phi(\alpha_t \mathbf{x}_0+\sigma_t\mathbf{\epsilon})-\epsilon\rVert _2^2\right],
    \label{eq:diffusion}
\end{equation}
where $w(t)$ is a time-dependent weighting function. After training, we have $p_t \approx q_t$; thus, we can draw samples from $p_0\approx q_0$. One of the most important applications is text-to-image generation \cite{rombach2022high, ramesh2022hierarchical}, where the noise prediction model $\epsilon_\phi(\mathbf{x}_t,t,y)$ is conditioned on a text prompt $y$.

\textbf{Text-to-3D generation by score distillation sampling (SDS) \cite{poole2022dreamfusion}.} SDS is widely used in text-to-3D generation \cite{lin2023magic3d, wang2024prolificdreamer, liang2024luciddreamer}, which lift 2D information upto 3D NeRF by distilling pre-trained diffusion models. Given a pre-trained text-to-image diffusion model $p_t(\mathbf{x}_t|y)$ with the noise prediction network $\epsilon_\phi(\mathbf{x}_t,t,y)$, SDS optimizes a single NeRF with parameter $\theta$. Given a camera viewport $\mathbf{p}$, a prompt $y$ and a differentiable rendering mapping $\mathbf{g}(\theta,\mathbf{p})$, SDS optimize the NeRF $\theta$ by minimizing:
\begin{equation}
    \begin{aligned}
        \mathcal{L}_{\text{SDS}}(\theta)=\mathbb{E}_{t, \mathbf{p}}\left[\frac{\sigma_t}{\alpha_t}w(t)D_{\text{KL}}(q_t^\theta(\mathbf{x}_t|c)\lVert p_t(\mathbf{x}_t|y^c))\right],
    \end{aligned}
    \label{eq:sds}
\end{equation}
where $\mathbf{x}_t=\alpha_t\mathbf{g}(\theta,\mathbf{p})+\sigma_t\epsilon$. Its gradients are approximated by:
\begin{equation}
    \begin{aligned}
        \nabla_\theta \mathcal{L}_{\text{SDS}}=\mathbb{E}_{t,\mathbf{p}}\left[w(t)\left(\epsilon_\phi(\mathbf{x}_t,t,y)-\epsilon\right)\frac{\partial \mathbf{g}(\theta, \mathbf{p})}{\partial \theta}\right].
    \end{aligned}
\end{equation}

\section{Proposed Method}

\begin{figure*}[htb]
    \centering
    \includegraphics[width=\linewidth]{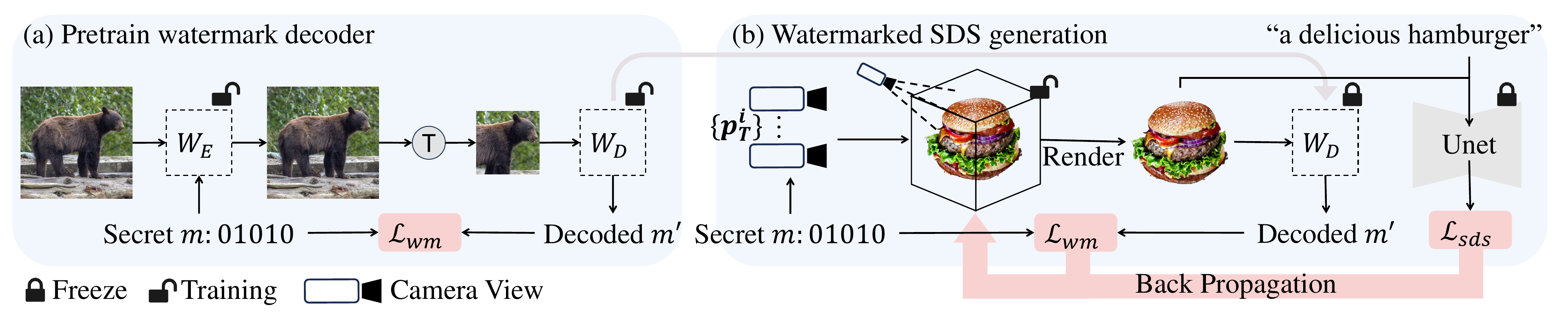}
    \caption{Overview. (a) We first pre-train a watermark encoder $W_E$ to embed a watermark into images and a watermark decoder $W_D$ to decode a watermark from images. (b) We generate trigger viewports $\{\mathbf{p}_T^i\}$ from the given secret message $m$ and optimize a NeRF such that the secret message can be decoded from images rendered from arbitrary trigger viewport $\mathbf{p}_T^i$.}
    \label{fig:pipeline}
\end{figure*}

\tool watermarks generation process of neural radiance fields (NeRF) by injecting backdoors during score distillation sampling (SDS). The message can be extracted from the rendered image of trigger viewports through a fixed watermark decoder. Our method is conducted in two phases. First, we pre-train the watermark decoder $W_D$. Then, we inject backdoors into a high-resolution NeRF during SDS optimization, such that images rendered from the trigger viewports yield a secret message.

\subsection{Pre-train the watermark decoder}

Different from CopyRNeRF \cite{luo2023copyrnerf}, which trains a separate watermark decoder for each watermarked NeRF, \tool employs a unique watermark decoder. This allows the NeRFs generated by our method to be verified using this unique decoder.

\textbf{Building watermark decoder training pipeline.} For simplicity, we use HiDDeN \cite{zhu2018hidden} as our $W_D$ architecture, a well-established image watermarking pipeline. It optimizes watermark encoder $W_E$ and watermark decoder $W_D$ for signature embedding and extraction. The encoder $W_E$ takes a cover image $x_o\in \mathbb{R}^{H\times W \times 3}$ and a $k$-bit message $m\in \{0,1\}^k$ as input and outputs a watermarked image $x_w\in \mathbb{R}^{H\times W \times 3}$. The decoder takes watermarked image $x_w$ as input and outputs a predicted secret message $m'$. The extracted message $m'$ is restricted to $[0,1]$ by utilizing a sigmoid function. The message loss is calculated with Binary Cross Entropy (BCE) between $m$ and sigmoid $sg(m')$:
\begin{equation}
    \mathcal{L}_{msg}=-\sum_{i=0}^{L-1}m_i\cdot \log sg(m'_i)+(1-m_i)\cdot \log (1-sg(m'_i)).
\end{equation}
The $W_E$ is discarded in the later phase since only $W_D$ serves our purpose.

Original HiDDeN architecture combines message loss and perceptual loss to optimize both $W_E$ and $W_D$. However, since $W_E$ is discarded and the perceptual loss is not needed, we follow the tradition \cite{fernandez2023stable, jang2024waterf} to optimize $W_E$ and $W_D$ by message loss only. We find that when the decoder receives a vanilla-rendered image, there is a bias between the extracted message bits. Thus, after training the decoder, we conduct PCA whitening to a linear decoder layer to remove the bias without reducing the extraction ability.

\textbf{Transformation layers.} For robustness consideration, a transformation layer is added between $W_E$ and $W_D$, which applies additional distortions to $x_w$, such as Gaussian blur, to make the decoder $W_D$ robust to multiple attacks. During training, it takes in a watermarked image $x_w$ produced by image encoder $W_E$ and outputs a noised version of the watermarked image, which will be further fed to the decoder $W_D$. This transformation layer is made of cropping, resizing, rotation and identity. Within each iteration, one type of transformation is chosen randomly for image editing. In detail, we add random cropping with parameters 0.3 and 0.7, resizing with parameters 0.3 and 0.7 and rotation with a degree range from $-\pi/6$ to $\pi/6$.

\subsection{Dreamark}

Inspired from existing black-box model watermarking \cite{adi2018turning, zhang2018protecting, jia2021entangled, le2020adversarial, chen2019blackmarks, szyller2021dawn}, where they root backdoors in a deep neural network to achieve DNN watermarking, we watermark NeRF by rooting backdoors during SDS optimization. Formally, given a NeRF with parameter $\theta$, a prompt $y$, a secret message $m$, we aim to optimize the NeRF such that the message $m$ can be decoded by $W_D$ from the image $\mathbf{g}(\theta, \mathbf{p}_T)$ rendered from arbitrary trigger viewport $\mathbf{p}_T$.

\textbf{Generate Trigger Viewports.} We wish to generate a set of trigger viewports $\{\mathbf{p}_T^i\}_{i=1}^N$ from the secret message $m$ such that the watermark verifier does not need to keep a replica of the trigger viewport set. Besides, different messages should generate different viewports because a constant trigger viewport set is easy to predict, leading to potential vulnerability. We use a pseudo-random number generator (PRNG) to generate the $m$-dependent trigger viewport set $\{\mathbf{p}_T^i\}_{i=1}^N$ as shown in Algorithm \ref{alg:trigger-viewport-generation}.

\begin{algorithm}
\caption{Trigger Viewport Generation}
\label{alg:trigger-viewport-generation}
\hspace*{\algorithmicindent} \textbf{Input:} Secret message $m$\\
\hspace*{\algorithmicindent} \textbf{Output:} $m$-dependent Trigger viewport $\{\mathbf{p}_T^i\}_{i=1}^N$
\begin{algorithmic}[1]
\State $seed \gets \text{SHA256}(m)$ \Comment{SHA-256 hash of the message}
\State $\text{Initialize random generator with } seed$
\State $\{\mathbf{p}_T^i\}_{i=1}^N \gets \text{Generate $N$ viewports with initialized generator}$
\State \Return $\{\mathbf{p}_T^i\}_{i=1}^N$
\end{algorithmic}
\end{algorithm}

\textbf{Choosing Trigger Embedding Media.} After generating $m$-dependent trigger viewport set $\{\mathbf{p}_T^i\}_{i=1}^N$, the question is how to choose suitable cover media to hide secret message $m$, such that $m$ can be decoded from the image rendered from arbitrary trigger viewport $\mathbf{p}_T^i$. In the text-to-NeRF generation \cite{wang2024prolificdreamer, liang2024luciddreamer, lin2023magic3d}, NeRF sample a set of points and obtains their colors $\mathbf{c}$ and geometry values $\sigma$ through two $\mathrm{MLP}$s: $f_\sigma, f_\mathbf{c}$ (Eq. \eqref{eq:sigma}, \eqref{eq:color}). For brevity, we can view the way NeRF computes point colors as $\mathbf{c}=\mathrm{MLP}(x,d)$. However, CopyRNeRF must learn a separate message feature field to get the point color $\mathbf{c}=\mathrm{MLP}(x,d,m)$. When integrating CopyRNeRF with an arbitrary text-to-NeRF pipeline, modifications to the NeRF structure are necessary to accommodate CopyRNeRF.

If we expect no additional changes to the NeRF structure, there are two options to hide triggers in the geometric mapping $f_\sigma$ or the color mapping $f_\mathbf{c}$. In practice, we find that the generated NeRF has low rendering quality once we incorporate geometric mapping $f_\sigma$ into backdooring. This may be because changing the geometric density of a sampled point will affect its color rendered from all viewing directions (Eq.(\ref{eq:color})). Therefore, we decide to backdoor in color mapping $f_\mathbf{c}$.

\textbf{Two-stage Trigger Embedding.} To backdoor in color mapping $f_\mathbf{c}$ only, we divide SDS optimization into two stages. In the first stage, we optimize a high-resolution NeRF (\eg, 512) by SDS (Eq.(\ref{eq:sds})) with joint optimization of both $f_\mathbf{c}$ and $f_\sigma$. The aim of the first stage is to generate scenes with complex geometry. In the second stage, we freeze $f_\sigma$ to fine-tune $f_\mathbf{c}$ by the following combined loss to conceal watermarks in trigger viewports $p_T$:
\begin{equation}
    \begin{aligned}
        \mathcal{L}_{\text{comb}}(\theta)=\mathcal{L}_{\text{sds}}+\mathbb{E}_{\mathbf{p}_T^i}\left[\mathrm{BCE}(W_D(\mathbf{g}(\theta, \mathbf{p}_T^i)), m)\right].
    \end{aligned}
    \label{eq:comb}
\end{equation}
Note that equation \ref{eq:comb} optimizes $\theta$ by SDS across arbitrary viewports $\mathbf{p}$, and $\mathrm{BCE}$ across trigger viewports $\mathbf{p}_T^i$.

\textbf{Watermark Extraction.} Given a suspicious NeRF $\mathbf{g}(\theta, \mathbf{p})$, the NeRF creator can first generate the trigger viewport set $\{\mathbf{p}_T^i\}_{i=1}^N$ following Algorithm \ref{alg:trigger-viewport-generation} based on his secret message $m$. Then the decoded message $m'$ can be extracted from the image rendered from arbitrary trigger viewport $\mathbf{p}_T\in \{\mathbf{p}_T^i\}_{i=1}^N$. The ownership can be verified by evaluating the bitwise accuracy between $m'$ and $m$.

\subsection{Implementation Details}
\begin{figure*}
    \centering
    \includegraphics[width=\linewidth]{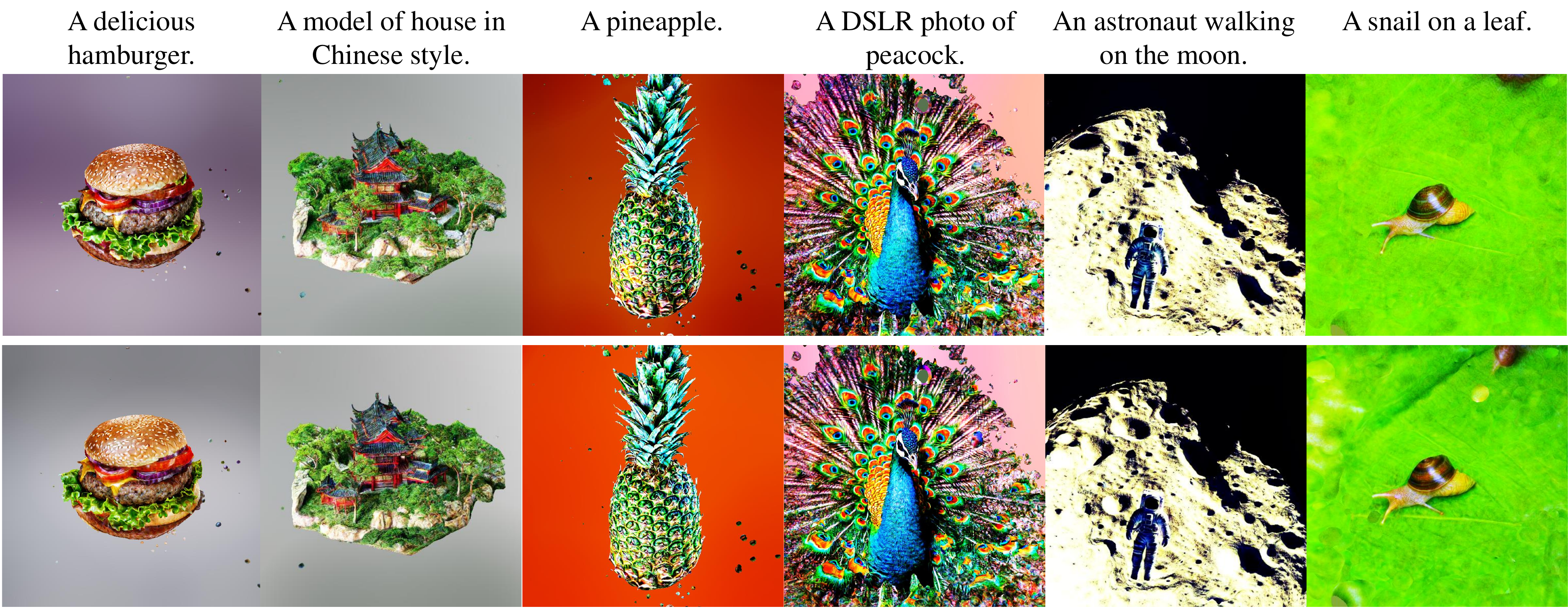}
    \caption{Images rendered from trigger viewports. Top: generated non-watermarked NeRF. Bottom: Watermarked NeRF generated by \tool. We aim to show that generated NeRF has the same visual quality as the non-watermarked NeRF instead of showing they are perceptually the same since there is no such the ``original NeRF" in the generation context.}
    \label{fig:visual-results}
\end{figure*}
\textbf{Pretrained Watermark Extractor.} We pretrain the watermark encoder $W_E$ and extractor $W_D$ using COCO \cite{lin2014microsoftcoco} dataset. We build $W_E$ with four-layer MLPs and $W_D$ with eight-layer MLPs, with all intermediate channels set to 64. During pretraining, the input image resolution is set to $256\times 256$, and the output message length is set to 48 to satisfy the capacity requirements of downstream watermarking tasks. We use Lamb \cite{you2019lamb} and CosineLRScheduler to schedule the learning rate, which decays to $1\times 10^{-6}$. This process is done in 500 epochs.

\textbf{NeRF.} We choose Instant NGP \cite{muller2022instant} for efficient high-resolution (\eg, up to 512) rendering. Given input coordinate $\mathbf{x}$, we use a 16-level progressive grid for input encoding with the coarsest and finest grid resolution set to 16 and 2048, respectively. The encoded input is further fed into $f_\mathbf{c}$ and $f_\sigma$, which are both built with one-layer MLPs with 64 channels, to predict the color $c_j$ and density $\sigma_j$ of input $\mathbf{x}$. We follow the object-centric initialization used in Magic3D \cite{lin2023magic3d} to initialze density for NeRF as $\sigma_{\text{init}}(\mathbf{x})=\lambda_\sigma(1-\frac{\lVert \mathbf{x}\rVert_2}{r})$, where $\lambda_\sigma=10$ is the density strength, $r=0.5$ is the density radius and $\mathbf{x}$ is the coordinate. We use Adam optimizer with learning rate $10^{-3}$ to optimize NeRF in both stages. The guidance model is Stable Diffusion \cite{rombach2022high} with the guidance scale set to 100. During SDS optimization, we sample time $t\sim \mathcal{U}(0.02, 0.98)$ in each iteration. We jointly optimize $f_\mathbf{c}$ and $f_\sigma$ for 40000 iterations in stage one and fine-tune $f_\mathbf{c}$ only for 30000 iterations in stage two.
\section{Experiment}

%AUC plot
%visual result after attack

% 比较message length
% trade-off between quality and acc
% robustness comparison
% performance under trigger size variation
% model-level robustness

% This section shows the generation quality, robustness, and capacity of \tool. We also compare \tool with post-generation watermarking algorithms: CopyRNeRF \cite{luo2023copyrnerf} and WateRF \cite{jang2024waterf}.

\subsection{Experiment Setup}

\begin{figure}[b]
    \centering
    \includegraphics[width=\linewidth]{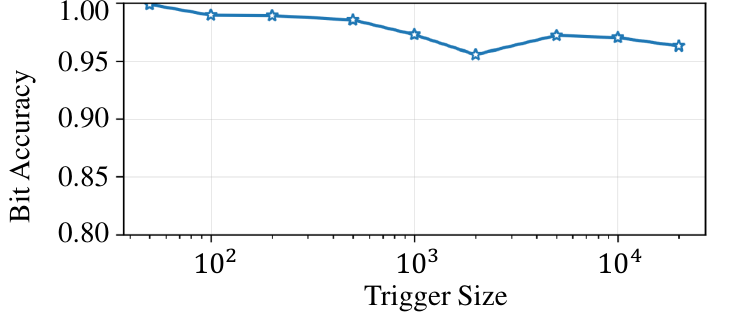}
    \caption{Effect under varied trigger size. Bit accuracy is not significantly affected by trigger size.}
    \label{fig:trigger-size}
\end{figure}

We select 16-bit secret messages and $N=1000$ trigger viewports in our experiment unless explicitly mentioned. To evaluate \tool, we use 100 different prompts to generate 100 watermarked scenes. Note that the scale of our experiment far exceeds that of prior works where they only evaluate Blender \cite{mildenhall2021nerf} and LLFF \cite{mildenhall2019local} dataset, with each dataset only containing eight scenes. All our experiments are conducted on Ubuntu 22.04 with an Intel Xeon Gold 5318Y CPU and an NVIDIA A100.

\textbf{Evaluation Metrics.} Two key evaluations for watermarking algorithms are invisibility and robustness. We evaluate robustness using bit accuracy under various image distortions such as Gaussian Noise, Rotation, Scaling, Gaussian Blur, Crop, and Brightness. For invisibility evaluation, different from the previous \textit{post-generation} watermarking algorithm, there is no such the ``original NeRF". Hence, the typical evaluation metric, the Peak-Signal-to-Noise Ratio (PSNR), is not applicable to evaluate our method. We follow prior 2D \textit{during-generation} watermarking algorithm \cite{yang2024gaussian, wen2024tree} where they use CLIP-Score \cite{radford2021clip} to evaluate the bias introduced by the watermarking algorithm.

\textbf{Baselines.} Since \tool is the first \textit{during-generation} watermarking method to watermark the generated NeRF, and existing NeRF watermarking CopyRNeRF \cite{luo2023copyrnerf} and WateNeRF \cite{jang2024waterf} can only embed watermark after NeRF generation. We select CopyRNeRF and WateNeRF as two \textit{post-generation} baselines, \ie we first generate NeRF with SDS only, then watermark the generated NeRF with CopyRNeRF and WateNeRF.

\begin{table}[tb]
    \centering
    \begin{tabular}{l|cccc}
    \toprule
         \multirow{2}{*}{Method} & \multicolumn{4}{c}{Bit Accuracy (\%)}\\
         & 8 bits & 16 bits & 32 bits & 48 bits\\
    \midrule
    \textit{Post Generation} &  &  &  & \\
    SDS+CopyRNeRF & 100\% & 91.16\% & 78.08\% & 60.06\%\\
    SDS+WateRF & 100\% & 94.24\% & 86.81\% & 70.43\%\\
    \midrule
    \textit{During Generation} &  &  &  & \\
    \tool & 100\% & 98.93\% & 82.59\% & 71.91\%\\
    \bottomrule
    \end{tabular}
    \caption{Bit accuracy under different bit length settings.}
    \label{tab:capacity}
\end{table}

\begin{table}[tb]
    \centering
    \begin{tabular}{l|ccc}
    \toprule
         Method & Bit Accuracy & CLIP/16 & CLIP/32\\
    \midrule
    None & N/A & 0.3156 & 0.2859\\
    \midrule
    \textit{Post Generation} &  &  & \\
    SDS+CopyRNeRF & 91.16\% & 0.3152 & 0.2831\\
    SDS+WateRF & 94.24\% & 0.3164 & 0.2823\\
    \midrule
    \textit{During Generation} &  &  & \\
    \tool & 98.93\% & 0.3218 & 0.2943\\
    \bottomrule
    \end{tabular}
    \caption{Bit accuracy and CLIP Score comparison with post-generation methods. ``None" reports the performance when no watermark is applied, so bit accuracy is not applicable.}
    \label{tab:acc}
\end{table}

\begin{table}[t]
    \centering
    \begin{tabular}{lcccc}
    \toprule
        & GN. & Rot. & Sca. & Crop\\
    \midrule
       Without $T$ & 0.64 & 0.57 & 0.56 & 0.89\\
       With $T$ & 0.93 & 0.84 & 0.92 & 0.91\\
    \bottomrule
    \end{tabular}
    \caption{Bit accuracy by removing transformation layer $T$.}
    \label{tab:transform_layer}
\end{table}
\begin{table}[t]
    \centering
    \begin{tabular}{cccc}
    \toprule
         &  CLIP/16 & CLIP/32 & Bit Acc\\
    \midrule
        Both $f_\mathbf{c}$ and $f_\sigma$ & 0.2308 & 0.2241 & 60.43\%\\
        $f_\mathbf{c}$ only & 0.3218 & 0.2943 & 98.93\%\\
    \bottomrule
    \end{tabular}
    \caption{Watermark by optimizing both $f_\mathbf{c}$ and $f_\sigma$ or $f_\mathbf{c}$ only.}
    \label{tab:geometric_color}
\end{table}

% \begin{figure*}[t]
%     \centering
%     \includegraphics[width=\linewidth]{figures/visual_attack.pdf}
%     \caption{Visualize image-level and model-level attacks. For image-level attacks, we use Gaussian Noise (v=0.1), 40\% center
% cropping and brightness (factor=2.0). For model-level attacks, we prune the smallest 12\% of the network parameters, which is
% evaluated on their $\ell_1$ norms, and perform fine-tuning attacks for 500 epochs.}
%     \label{fig:visual-attack}
% \end{figure*}

\subsection{Performance of Dreamark}

\textbf{Capacity.} We evaluate bit accuracy across 8, 16, 32, and 48-bit secret messages. Table \ref{tab:capacity} shows that all watermarking methods have a trade-off between bit accuracy and capacity. As a during-generation watermarking method, \tool shows relatively high accuracy on 8, 16, and 48-bit settings compared to post-generation watermarking methods, such as SDS+CopyRNeRF and SDS+WateRF. For example, in 16-bit settings \tool achieves 98.93\% accuracy while SDS+CopyRNeRF and SDS+WateRF achieve 91.16\% and 94.24\% accuracy, respectively.

\textbf{Generation quality.} We report CLIP/16 evaluated by \textit{clip-ViT-B-16} and CLIP/32 evaluated by \textit{clip-ViT-B-32} to indicate the generation quality of the watermarked images. For each scene, its CLIP-Score is averaged among all viewports $\mathbf{p}$, and bit accuracy is averaged among all $N$ trigger viewport $\{\mathbf{p}_T^i\}_{i=1}^N$. All prior watermarking works \cite{jang2024waterf, luo2023copyrnerf, zhu2024rethinking, zhu2018hidden} show a trade-off between bit accuracy and watermarked content quality. However, Table \ref{tab:acc} shows \tool achieves superior performance in both bit accuracy and generation quality when compared to post-generation methods. Notably, the CLIP Score is 0.3156/0.2859 when no watermark is applied, while \tool achieves 0.3218/0.2943 CLIP Score. This indicates that during-generation watermark embedding of \tool does not harm the generation quality. We also provide visual results of generated NeRF in Figure \ref{fig:visual-results}.

\begin{figure}[!hb]
    \centering
    \includegraphics[width=\linewidth]{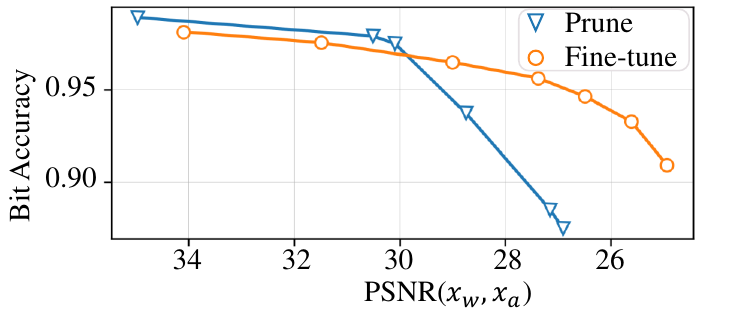}
    \caption{Robustness against model-level attacks. $x_w,x_a$ are images rendered from watermarked and attacked NeRF.}
    \label{fig:model-level-attack}
\end{figure}

\textbf{Size of trigger viewports.} We evaluate bit accuracy under different numbers of trigger sizes $N$. As is shown in Figure \ref{fig:trigger-size}, we surprisingly find that bit accuracy is not significantly affected by trigger size. For example, bit accuracy reaches 99.88\% when $N=50$ and only drops to 96.27\% when $N$ increases to $20000$.

% \textbf{Trade-off between generation quality and robustness.} We try to maximize the generation quality or the robustness of the watermark by varying the weight $\lambda$ in Equation \ref{eq:comb}. A higher $\lambda$ leads to a decoded message closer to the target secret message. 

% \begin{table}[t]
%     \centering
%     \begin{tabular}{lccccc}
%     \toprule
%         $\lambda$ & 0.1 & 1 & 5 & 10 & 100\\
%     \midrule
%         CLIP/16 & 0.3211 & 0.3218 & 0.3377 &  & \\
%         CLIP/32 & 0.2877 & 0.2943 & 0.3064 &  & \\
%         Bit Acc. & 58.40\% & 98.93\% & 99.96\% &  & \\
%     \bottomrule
%     \end{tabular}
%     \caption{Performance under varied $\lambda$.}
%     \label{tab:trade-off}
% \end{table}

\subsection{Ablation Study}

Can we remove the transformation layer $T$ during watermark decoder $W_D$ pre-training? Table \ref{tab:transform_layer} shows that the robustness significantly drops when the transformation layer is removed.

Can we hide watermarks by jointly optimizing color mapping $f_\mathbf{c}$ and geometric mapping $f_\sigma$? Table \ref{tab:geometric_color} shows that watermark, by optimizing both color mapping and geometric mapping, has lower performance on both bit accuracy and CLIP Score. This may be because, within one iteration, only one single direction of point color is supervised, while changing its point density will affect its color in all directions, significantly increasing the difficulty of convergence.

\section{Attacks on Dreamark's Watermarks}
\begin{table*}[htb]
    \centering
    \begin{tabular}{l|ccccccc}
    \toprule
         \multirow{2}{*}{} & \multicolumn{7}{c}{Bit Accuracy (\%)}\\
         & No Distortion & \makecell{Gaussian Noise\\(v=0.1)} & \makecell{Rotation\\($\pm \pi/6$)} & \makecell{Scaling\\(25\%)} & \makecell{Gaussian Blur\\(deviation=0.1)} & \makecell{Crop\\(40\%)} & \makecell{Brightness\\(2.0)}\\
    \midrule
    \textit{Post Generation} &  &  &  &  &  &  & \\
    SDS+CopyRNeRF & 91.16\% & 90.04\% & 88.13\% & 89.33\% & 90.06\% & N/A & N/A\\
    SDS+WateRF & 94.24\% & 94.06\% & 85.02\% & 91.35\% & 94.12\% & 95.40\% & 90.91\%\\
    \midrule
    \textit{During Generation} &  &  &  &  &  &  & \\
    \tool & 98.93\% & 93.75\% & 84.51\% & 92.40\% & 98.93\% & 91.49\% & 91.23\%\\
    \bottomrule
    \end{tabular}
    \caption{Robustness under multiple image transformations compared with post-generation-based methods.}
    \label{tab:rob}
\end{table*}

% \begin{table*}[htb]
%     \centering
%     \begin{tabular}{l|cccccc}
%     \toprule
%          \multirow{2}{*}{} & \multicolumn{6}{c}{Bit Accuracy (\%)}\\
%          & No Distortion & \makecell{Gaussian Noise\\(v=0.1)} &  \makecell{Scaling\\(25\%)} & \makecell{Gaussian Blur\\(deviation=0.1)} & \makecell{Crop\\(40\%)} & \makecell{Brightness\\(2.0)}\\
%     \midrule
%     \textit{Post Generation} &  &  &  &  &  & \\
%     SDS+CopyRNeRF & 91.16\% & 90.04\% & 89.33\% & 90.06\% & N/A & N/A\\
%     SDS+WateRF & 94.24\% & 94.06\% & 91.35\% & 94.12\% & 95.40\% & 90.91\%\\
%     \midrule
%     \textit{During Generation} &   &  &  &  &  & \\
%     \tool & 98.93\% & 93.75\% & 92.40\% & 98.93\% & 91.49\% & 91.23\%\\
%     \bottomrule
%     \end{tabular}
%     \caption{Robustness under multiple image transformations compared with post-generation-based methods.}
%     \label{tab:rob}
% \end{table*}

This section aims to examine the robustness of the watermark against various attacks. We first consider image-level attack, which performs arbitrary image transformations and is typical for many NeRF watermarking methods \cite{luo2023copyrnerf, jang2024waterf}. We then consider model-level attacks such as fine-tuning and pruning since the generated NeRF could be made public; in this case, the attacker will have white-box access to the generated NeRF. Besides, model-level attacks are commonly evaluated in model watermarking methods \cite{adi2018turning, zhang2018protecting, jia2021entangled, le2020adversarial, chen2019blackmarks, szyller2021dawn}.

\subsection{Robustness against image-level attacks}

We evaluate the robustness of the watermark against different image transformations before rendered images are fed into the watermark decoder. We consider Gaussian Noise (v=0.1), rotation ($\pm \pi/6$), Scaling ($25\%$), Gaussian Blur (deviation=0.1), Crop (40\%) and Brightness (2.0). Bit accuracy is averaged on all transformed images rendered from trigger viewports. Table \ref{tab:rob} shows \tool is robust against previously mentioned image-level attacks. The bit accuracy is always above 90\% except for rotation. Note that the robustness is achieved without the need for transformations during the \tool optimization phase: it is attributed to the watermark decoder. If the watermark decoder is trained to withstand arbitrary transformation, the generated NeRF subsequently learns to contain watermarks that maintain robustness throughout the \tool optimization.

\subsection{Robustness against model-level attacks}

This subsection considers the scenario when an attacker gets full access to the generated NeRF model and aims to remove the embedded watermark without degrading its visual quality. We denote $x_w$ as images rendered from watermarked NeRF and $x_a$ as images rendered from attacked NeRF. We use $\mathrm{PSNR}(x_w,x_a)=-10\cdot\log_{10}(\mathrm{MSE}(x_w,x_a))$, for $x_a,x_w\in [0,1]^{c\times h\times w}$ to evaluate distortion made by attacks.

\textbf{Model Fine-tuning.} Since our method uses a unified watermark decoder $W_D$ to decode the secret message, we consider two scenarios of fine-tuning attack. One assumes that the attacker has full access to the watermark decoder $W_D$, and the other assumes that the attacker has no access to $W_D$. Besides, we assume the attacker has no prior knowledge of the secret message $m$. In this case, the attacker cannot reproduce trigger viewports since trigger viewports are only related to $m$. For the first scenario, when the attacker has full access to $W_D$, the attacker can use an adversarial attack to partially remove the watermark by minimizing the $\mathrm{BCE}$ loss between the extracted message and a random binary message sampled beforehand:
\begin{equation}
    \begin{aligned}
        &\mathcal{L}_{\text{fine-tune}}(\theta')=\\
        &\mathbb{E}_{\mathbf{p}}[\rVert\mathbf{g}(\theta',\mathbf{p})-\mathbf{g}(\theta, \mathbf{p})\lVert_2^2
        +\mathrm{BCE}(W_D(\mathbf{g}(\theta',\mathbf{p})),m')],
    \end{aligned}
\end{equation}
where $\theta$ is the fixed parameter of watermarked NeRF, $\theta'$ is the parameter of NeRF to be fine-tuned, $m'$ is a random binary message different from $m$. As shown in Fig. \ref{fig:model-level-attack}, even if PSNR drops below 26dB, bit accuracy is still above 90\%. For the second scenario, when the attacker has no access to $W_D$, the attacker cannot produce the adversarial attack to remove the watermark.

\textbf{Model Pruning.} Model pruning is widely used in model compression since it can reduce the storage and computation cost of DNNs. However, pruning will affect not only the size and operation speed of the model but also the accuracy of the watermark and the visual quality of NeRF. A higher pruning rate gives lower watermark accuracy and lower visual quality. In practice, we vary the pruning rate and, at the same time, evaluate $\mathrm{PSNR}(x_a, x_w)$ and bit accuracy on $x_a$. The pruning attack with pruning rate $a$\% means setting the smallest $a$\% of network parameters to zero, where the size of the network parameter is evaluated by its $\ell_1$ norm. Fig. \ref{fig:model-level-attack} shows that our method is robust against pruning attack since we still have $\sim$88\% accuracy when PSNR is below 27dB, while 27dB PSNR means relatively high distortion has been made in image watermarking context \cite{zhu2018hidden}.
\section{Conclusion}

In this work, we propose a \textit{during-generation} text-to-3D watermarking method, \tool, which eliminates the delay between the generation phase and the watermarking phase: the watermark can be verified on the generated NeRF once the generation is finished. Inspired by the black-box model watermarking method, \tool watermarks NeRF by injecting backdoors into NeRF such that a secret message can be extracted from images rendered from arbitrary trigger viewport. Extensive experiments show that our method will not degrade generation quality and maintain robustness against image-level and model-level attacks.

\section*{Acknowledgments}
This work was supported in part by National Key R\&D Program of China under Grant 2021YFF0900300, in part by Guangdong Key Program under Grant 2021QN02X166, and in part by Research Institute of Trustworthy Autonomous Systems under Grant C211153201. Any opinions, findings, and conclusions or recommendations expressed in this material are those of the author(s) and do not necessarily reflect the views of the funding parties.

\bibliography{aaai25}

\end{document}